\documentstyle[preprint,aps]{revtex}
\newcommand{\be}{\begin{equation}}
\newcommand{\ee}{\end{equation}}
\newcommand{\bea}{\begin{eqnarray}}
\newcommand{\eea}{\end{eqnarray}}

\begin{document}
\preprint{SU-ITP-96-38, SLAC-PUB-7262}
\title{Fixed Scalars and Suppression of Hawking Evaporation }
\author{\bf Barak Kol}
\address{ Department of Physics, Stanford University, Stanford, CA
94305, USA}
\address{E-mail: barak@leland.stanford.edu}
\author{\bf  Arvind Rajaraman}
\address{Stanford Linear Accelerator Center,\\
    Stanford University, Stanford, California 94309 USA}
\address{E-mail: arvindra@dormouse.stanford.edu}
\maketitle
\begin{abstract}
For an extreme charged black hole some scalars take on a fixed value at the
horizon determined by the charges alone. We call them fixed scalars. We find
the  absorption cross section for a  low frequency wave of  a fixed scalar to
be proportional to the square of the frequency. This implies a strong
suppression of the Hawking radiation near
extremality. We compute the coefficient of proportionality for a specific
model.
\end{abstract}

\newpage
\bigskip
\rightline{bicurim - }
\rightline{first fruits}
\section {Introduction}

Consider an extreme charged black hole. As a specific example, we have in mind
a 4 dimensional N=4 supergravity model,
\cite{KLOPP}, that will be presented in the next section. Ferrara, Kallosh and
Strominger \cite{FKS}, \cite{FK} found that some scalars reach a fixed value at
the
horizon $\phi_H = \phi_{fix} (q)$. $\phi_{fix}$ depends only on the charges,
$q$, and is independent of the scalar background (the value of the scalars at
infinity) $\phi_{\infty}$. Other scalars which couple only to gravity will be
called
indifferent scalars. To determine which scalars are fixed for a given extreme
charged black hole, the charges of the black hole must be known. (In the case
we consider, there is simply one fixed scalar).

Previous consideration of the fixed scalars led \cite{FirstLaw} to the addition
of an extra term to the first law of black hole dynamics : $dM = ...-\Sigma
d\phi_\infty$,  where M is the mass and $\Sigma$ is the scalar charge. Thereby
the scalar charge was identified to be the thermodynamic conjugate of the
scalar background.

For a fixed scalar a change in the value of the field at infinity does not
change the value at the horizon, $\phi_{fix}$. Consider scattering a long
wavelength scalar wave off the black hole. Such a wave causes a slow change of
the scalar background. An observer located near the horizon would hardly feel
the presence of the wave since the field in the vicinity of the horizon is
protected from changes. One would therefore expect a low absorption cross
section for this scalar.  On the other hand, computations of cross
sections for indifferent scalars in the long wavelength limit were done
starting from the mid 70's (for example \cite{Unruh}, \cite{GWG}), and it is
known that the cross section is of the order of the area of the horizon. For a
fixed scalar we expect to find a novel effect : the cross section vanishes at
long wavelengths.

The vanishing of the cross section has several implications : first, it is
surprising to find some sort of radiation which is not absorbed by a black hole
. Second, by
definition, the Hawking radiation of a black hole balances with the absorption
from an environment with the same temperature. For a nearly extremal black hole
the cross section would be very small and thus the evaporation through the
emission of fixed scalars would be largely suppressed. Third, it would be
interesting to compare this result with the D-brane model of black holes
\cite{VfSr} \cite{ClMl} \cite{DsMt}.

To prove the vanishing of the cross section, we employ analytic and numerical
methods. In section II, we determine the linearized equation of motion for the
scalar, and find it to have an effective mass term near the horizon. In section
III, we look for a long wavelength solution of the wave equation in the black
hole background. We use the well known procedure of solution matching
(\cite{Unruh}). In the low frequency limit, the leading contribution (which
will turn out to vanish) comes from the s-wave. The boundary condition is that
at the horizon there is
only an ingoing wave, and the behavior at infinity gives the reflection
coefficient. Space is divided into three regions, in each region a different
approximation can be used giving three solutions that are finally matched on
overlap regions.

In order to gain further evidence, we performed numerical computations using
"Mathematica" \cite{math}
and observed a suppression of the cross section for a fixed scalar, compared to
an indifferent one. The numerical method complements the analytic one in
covering the frequency range : the analytic method describes only small
frequencies, and the numerical method requires the frequency to be not too
small.

The order of vanishing of the cross section must be (at least) $\omega ^2$.
That  is because the transmission amplitude T, satisfies $T(-\omega ) = T^*
(\omega )$ and thus $\sigma =  | T | ^2 $ is an even function $\sigma (-\omega
) = \sigma (\omega )$. We compute the numerical coefficient in a specific
setting.

\section{The Model}
We use a 4d N=4 supergravity model given by \cite{KLOPP}. The bosonic part of
the action is given by :
\begin{equation}
\int{d^4 x \sqrt{-g}  [ -R +2 \partial ^{\mu} \phi \partial _{\mu} \phi -
e^{-2\phi} F_{\mu \nu} F^{\mu \nu} - e^{2\phi}  \tilde{G}_{\mu \nu}
\tilde{G}^{\mu \nu} ]} \ .
\label{action}
\end{equation}
where g is the metric, R is the Ricci scalar, $\phi$ is a real scalar - the
dilaton, and $ F, \tilde{G}$ are two U(1) field strengths. The dilaton here is
a fixed scalar as shown in \cite{FK}.

The equations of motion are

\begin{eqnarray}
\partial _{\mu} (e^{-2\phi} F^{\mu \nu} )=0  \\
\partial _{\mu} (e^{2\phi} \tilde{G}^{\mu \nu} )=0  \\
\partial _{\mu} \partial ^{\mu} \phi - {1 \over 2} e^{-2\phi} F^2+ {1 \over 2}
e^{-2\phi} \tilde{G}^2 = 0\\
R_{\mu \nu} + 2 \partial _{\mu} \phi \partial _{\nu} \phi - e^{-2\phi} (2
F_{\mu\lambda}F_{\nu\delta}g^{\lambda\delta } - {1 \over 2} g_{\mu \nu} F^2 ) -
\nonumber \\
- e^{2\phi} (2 \tilde{G}_{\mu\lambda} \tilde{G}_{\nu\delta}g^{\lambda\delta } -
{1 \over 2} g_{\mu \nu} \tilde{G}^2 ) = 0 .
\label{eq-of-motion}
\end{eqnarray}

The extreme black hole solution, that we will use as background for the scalar
waves is
\begin{eqnarray}
ds^2 = e^{2U} dt^2 - e^{-2U} dr^2 - R^2 d^2 \Omega \nonumber \\
e^{2U} = (1-M/r)^2 \nonumber \\
R^2 = r^2 \nonumber \\
e^{2\phi} = 1 \nonumber \\
F = {Q \over r^2} dt \wedge dr \nonumber \\
\tilde{G} = {P \over r^2} dt \wedge dr
\label{soln}
\end{eqnarray}
with
 $Q = P = M/\sqrt{2} $.
This is an extreme Reissner-Nordstr\"{o}m solution with mass M. The
horizon is located at $r=M$.

To get the wave equation for small spherical perturbations we shall linearize
the
equations of motion around the above solution. Spherical perturbations of
the metric can be described  \cite{MTW} by
\begin{equation}
ds^2 = g_{tt}(t,r) dt^2 - g_{rr}(t,r) dr^2 -r^2 d^2 \Omega .
\end{equation}
In this coordinate system the area of a spherical shell is always $4\pi r^2$.
Gauss' law gives :
\begin{equation}
\label{Gauss}
4\pi r^2 e^{-2\phi} F^{rt} = \oint_{S^2_r}{e^{-2\phi} \star F} = 4\pi Q
\end{equation}
We denote $F^{rt} = E_F, \tilde{G}^{rt} = E_{\tilde{G}}$ - electric fields.
Then
\begin{equation}
F^2 = 2 E_F^2 g_{tt} g_{rr} = -2E_F^2 = -2 {Q^2 \over r^4}
\label{EF}
\end{equation}
Performing the variation of (\ref{Gauss}) we get \begin{equation}
\delta E_F = 2 \phi E_F
\label{dF}
\end{equation}
Similarly for $\tilde{G}$ :
\begin{equation}
\delta E_{\tilde{G}} = -  2 \phi E_{\tilde{G}} .
\end{equation}

The variation of the dilaton equation gives
\begin{eqnarray}
0 =\partial _{\mu} \partial ^{\mu} \phi + \phi (F^2 + \tilde{G}^2) - {1 \over
2} \phi \ \delta(F^2 - \tilde{G}^2) \nonumber \\
\delta(F^2 - \tilde{G}^2)  = \delta ( 2 E_{\tilde{G}}^2 g_{tt} g_{rr} - 2 E_F^2
g_{tt} g_{rr}) = \nonumber \\
= 4 ( E_{\tilde{G}} \delta E_{\tilde{G}} - E_F \delta E_F) g_{tt} g_{rr} + 2
(E_{\tilde{G}}^2 - E_F^2) \delta (g_{tt} g_{rr})
\end{eqnarray}
The second term vanishes because the solution we expand around satisfies
$E_{\tilde{G}}^2 = E_F^2$. Using the expression for $\delta E_F , \delta
E_{\tilde{G}}$ (\ref{dF}) we get :
\begin{equation}
\delta(F^2 - \tilde{G}^2) = -8 (E_{\tilde{G}}^2 + E_F^2).
\end{equation}
Substituting back, and using (\ref{EF})
\begin{eqnarray}
0 = \partial _{\mu} \partial ^{\mu} \phi -2 \phi (E_{\tilde{G}}^2 + E_F^2 ) +
4\phi (E_{\tilde{G}}^2 + E_F^2 ) = \partial _{\mu} \partial ^{\mu} \phi + 2
\phi (E_{\tilde{G}}^2 + E_F^2 )  \Rightarrow \nonumber \\
0 = \partial _{\mu} \partial ^{\mu} \phi + 2 {P^2 + Q^2 \over r^4} \phi=
\partial _{\mu} \partial ^{\mu} \phi + 2 {P^2 + Q^2 \over r^4} \phi
\label{4dEq}
\end{eqnarray}
Eq. (\ref{4dEq}) is the linearized equation for $\phi$. This equation describes
a scalar with a space dependent mass term
\begin{equation}
m_{\phi}^2 = 2 {M^2 \over r^4}
\end{equation}

To get the radial equation we separate variables
\begin{equation}
\phi = exp (-i\omega t)\ Y_{lm}\  \phi (r)
\end{equation}
Using the equation for the wave operator in curved space-time, we get
\begin{eqnarray}
0 = [ -{1 \over g_{tt}} \omega ^2 + {l(l+1) \over r^2} - {1 \over r^2
\sqrt{g_{tt} g_{rr}} } {d \over dr} r^2 \sqrt{{g_{tt} \over g_{rr}}} {d \over
dr} + {2 M^2 \over r^4} ] \phi \\
 =[ -(1-{M \over r})^{-2} \omega ^2 + {l(l+1) \over r^2} - {1 \over r^2} {d
\over dr} r^2(1-{M \over r})^2 {d \over dr} + {2 M^2 \over r^4} ] \phi
\end{eqnarray}

Choosing $l=0$, and setting the unit of length $M=1$, we finally have

\begin{equation}
0 = [ {d^2 \over dr^2} + {2 \over (r-1)} {d \over dr} + {r^4 \over (r-1)^4}
\omega ^2 - {2 \over r^2 (r-1)^2} ] \phi .
\end{equation}
This is the differential equation we will study. The horizon is located at
$r=1$.

\section{The Solution}
We will find an approximate solution by solving the equation in three regions,
and
matching the solutions. The three regions are found by examining which of the
two potential terms is dominant. The condition at the overlap region is
\be
{r^4 \over (r-1)^4}\omega^2 \simeq {2 \over r^2(r-1)^2}
\ee
which has the two solutions
\be
r-1 \simeq {\omega \over \sqrt{2}} \qquad r\simeq {(2)^{1 \over 4}
\over \sqrt{\omega}}
\ee

In the near horizon region $1 < r < 1+ {\omega \over \sqrt{2 }}$, we substitute
$z=\frac{z}{z-1}$ and drop terms which are higher order than $\frac{1}{z^2}$.
This yields the differential equation

\be
\left({d^2 \over dz^2} +  (1+ {4\over z})\omega^2-{2 \over z^2}
 \right) \phi =0
\ee
Rescaling $\zeta=\omega z$, we can write the equation as
\be
\left({d^2 \over d\zeta^2} + (1- {2\eta \over \zeta}-{2L(L+1) \over \zeta^2})
\right) (\phi)=0
\ee
with $\eta=-2\omega, L=1$, which has the solution in terms of Coulomb wave
functions \cite{AS}
\be
\phi = A F_1(\zeta) + B G_1(\zeta)
\ee
Since we must have ingoing waves at the horizon (i.e. large $\zeta$), we will
choose the solution
 \be
\phi_1 = i F_1(\zeta) +  G_1(\zeta)
\label{phi1}
\ee

The second region is defined by $1+ {\omega \over \sqrt{2 }} < r <
{(2)^{1 \over 4} \over \sqrt{\omega}}$. In this region we can drop the
term with $\omega^2$. The differential
equation becomes
\be
\left({d^2 \over dr^2} + {2 \over r-1}{d \over dr} -{2 \over r^2(r-1)^2}
 \right) \phi =0
\ee
which has the solution
\be
\phi_2= A \left(1-{1 \over r} \right) + B\left(1-{1 \over r} \right)^{-2}
\ee

The third region occurs for $r > {(2)^{1 \over 4} \over \sqrt{\omega}}$.
In this region we keep terms up to order $1/r$. The differential equation
becomes
\be
\left({d^2 \over dr^2} + {2 \over{r-1}}{d \over dr} + (1+{4 \over r})\omega^2
 \right) \phi =0
\ee
Scaling $\rho=\omega r$, we get the equation
\be
\left({d^2 \over d\rho^2} + (1- 2\eta/\rho) \right) (r\phi)=0
\ee
with $\eta=-2\omega$.
The solution can again be written in terms of Coulomb functions \cite{AS}
\be
r\phi_3= C F_0(\rho) + D G_0(\rho)
\ee

We now perform the matching across the various regions.

For small $\zeta$ the solution (\ref{phi1}) becomes \cite{AS}
\bea
\phi_1= iC_1(\eta) \zeta^2 +{1 \over 3C_1(\eta)\zeta}\\
=iC_1(\eta) \omega^2 z^2 +{1 \over 3C_1(\eta)\omega z}
\eea
where $C_1(\eta)= {1 \over 3} +$ (terms of order $\eta$).
The intermediate solution can be written as
\be
\phi_2= {A \over z} + Bz^2
\ee
Matching the two solutions, we find
\be
\frac{B}{A}= {i\omega^3 \over 3}
\ee

We now match the intermediate and far regions. We do this by considering an
asymptotic expansion of  $\phi_3$ for small $\rho$ and matching to an expansion
of $\phi_2$ for large $r$.

For small $\rho$, we have \cite{AS}
\bea
\phi_3 &=& {1 \over \rho}\left( C C_0(\eta) \rho + {1 \over C_0(\eta)} D
\right)\\
&=& C C_0(\eta) + {D \over C_0(\eta)\omega r}
\eea
where $C_0(\eta)=e^{\frac{\pi\eta}{2}}\|\Gamma(1+i\eta)\|$.

For large $r$, we have
\be
\phi_2= (A+B) +{1 \over r} (2B-A)
\ee

Matching the two solutions, we get
\be
{D \over C}=C_0(\eta)^2\omega (2B-A)/(A+B)
\ee

Using $C_0 = 1$, we can then evaluate the absorption coefficient
\be
A_\alpha=1-\|{C+iD \over C-iD}\|^2
\ee
which comes out to be
\be A_\alpha = 4 \omega^4 \ee

The corresponding s-wave cross section is then found to be
\be
\sigma_S={A_\alpha \pi \over \omega^2}= 4\pi\omega^2
\ee

Inserting the dependence on $M$, the s-wave cross section is
\be
\sigma_S=4\pi\omega^2 M^4
\ee
\section {Summary}

We have evaluated the absorption cross-section for a fixed scalar in an N=4
model, and
have shown that it vanishes for low frequencies. By detailed balance, this
implies a low
rate of Hawking evaporation for these scalars.

\begin{center}
{\bf ACKNOWLEDGEMENTS}
\end{center}

We thank L. Susskind and G.W.Gibbons.

 The work of B.K. is supported by NSF grant PHY-9219345. The work of A.R. is
supported
in part by the Department of Energy
under contract no. DE-AC03-76SF00515.

\end{document}